\documentclass[aps,nofootinbib,twocolumn,citeautoscript]{revtex4-2}
\usepackage{amsmath,amssymb}
\usepackage{enumerate}
\usepackage{xcolor}
\usepackage[colorlinks=true,citecolor=red,linkcolor=blue,linkbordercolor=green]{hyperref}
\usepackage{breqn}
\usepackage{setspace}
\usepackage{romannum}
\usepackage{enumerate}
\usepackage{graphicx}
\usepackage{float}

\date{\today}

\begin{document}
\title{Trade-off between Squashed Entanglement and Concurrence in Bipartite Quantum States}
\author{Kapil K. Sharma$^{*}$ and Suprabhat Sinha$^{\dagger}$  \\\vspace{0.4cm}
\textit{$^{*}$DY Patil International University,\\
Sect-29, Nigdi Pradhikaran, Akurdi,\\
Pune, Maharashtra-411044, India} \\
E-mail: $^{*}$iitbkapil@gmail.com \\\vspace{0.4cm}
\textit{$^\dagger$Department of Electronics,\\
West Bengal State University \\
Barasat, Kolkata-700126, India} \\ 
Email: $^\dagger$suprabhatsinha64@gmail.com}

\begin{abstract}
In this article we investigate the unitary dynamics of squashed entanglement and concurrence measures in Werner state and maximally entangled mixed states (MEMS) under two different Hamiltonians. The aim of the present study is two fold. The first part of the study deals with the dynamics under Heisenberg Hamiltonian and the second part deals under bi-linear bi-quadratic Hamiltonian which is the extension of first Hamiltonian. In both  the parts we investigate the dynamical trade off and balancing points for squashed entanglement and concurrence. During the study, we also found the results of entanglement sudden death (ESD) with Heisenberg Hamiltonian in Werner state under concurrence measure. In second part, we investigate the special result for bi-linear bi-quadratic Hamiltonian; it does not disturb squashed entanglement and concurrence in both the states and exhibit the robust character for both of the states. 
\end{abstract}
\maketitle

\section{Introduction}
Quantum information and computation is an emerging area which has potential impact on future quantum technologies\cite{nl}. This area has done tremendous progress in recent  years and currently many commercial companies have interest to explore the cloud services for Noisy Intermediate Scale Quantum Computers (NISQ)\cite{NISQ} by 2025. These quantum computers are also rephrased as Near Term Quantum Computers, which will be equipped with 50 to 100 qubits. Despite the fact that quantum information and computation is very emerging but on the other hand this area also have theoretical hindrances which are still unexplored\cite{pr}. A physical system incorporating the superposition, prolonged entanglement and minimum decoherence play an important role in quantum information. The scientific community always have keen interest to discover quantum materials\cite{qm} which can perform the quantum computation at room temperature. To the date, quantum computer function at very low temperature and dilute refrigerator is the primary requirement for these computers.   

Exploring the entanglement dynamics in quantum information needs to consider varieties of quantum states in the background; since the quantum states are the building blocks for quantum technologies incorporating quantum teleportation\cite{QT1,QT2,QT3}, quantum sensing\cite{QS},  quantum cryptography\cite{QC1,QC2} etc. Entanglement is very fragile phenomenon and very sensitive towards external agents; also it is very difficult to maintain it for a long time in a physical system. The experimental manifestation for prolonged entanglement in varieties of quantum states are on the way. On the theoretical side, it is also difficult to characterize the multipartite entanglement by following the quantum resource theory\cite{rth}. The dynamical study of quantum correlations and investigating their sustainability in varieties of quantum states is an important regime for quantum information. The various theoretical entanglement measures like Wootters concurrence\cite{CON} and quantum discord\cite{QD1,QD2,QD3} are tested in varieties of spin chains along with experimental manifestations. In this direction it is also important to proceed theoretical studies with other quantum correlations measures. 

In the current article we focus on the dynamical aspects of quantum correlations measures such as squashed entanglement\cite{sqe1} and concurrence\cite{CON} in bipartite Werner state\cite{WS} and maximally entanglement mixed states (MEMS)\cite{MEMS}. The study is carried out under two different Hamiltonians called as Heisenberg Hamiltonian\cite{H1} and bi-linear bi-quadratic Hamiltonian\cite{H2}. The bi-linear bi-quadratic Hamiltonian is the non linear version of Heisenberg Hamiltonian and both the Hamiltonians supports $SU(2)$ symmetry. The entanglement dynamics under Heisenberg Hamiltonian has been studied in varieties of spin chains configured in thermal and non-thermal conditions\cite{ss,te}. But in literature the study on bi-linear bi-quadratic Hamilton is very limited. 

Following the literature, the present work pursue the study under two different kinds of Hamiltonians and we have found an interesting property that bi-linear bi-quadratic Hamiltonian does not effect the quantum correlations in both the states (Werner and MEMS). In 2011, squashed entanglement\cite{sqe1,sqe2} attracted much attention of quantum community, but the simulations of squashed entanglement in larger Hilbert spaces are also missing to till date. So, in this direction the present study is totally new in this direction to the best of our knowledge. 

The outlines of the paper is sketched in \Romannum{8} sections. In section  \Romannum{2}, we present the brief introduction of Werner state and MEMS and characterization of quantum correlations with squashed entanglement and concurrence. Section \Romannum{3} highlights the Hamiltonians applied in the work and approach of unitary time evolution. Section \Romannum{4} is devoted to initial state preparation of quantum states used in the work. In section \Romannum{5}, we obtain  the mathematical functions of squashed entanglement and concurrence with time evolution under Heisenberg Hamiltonian. The dynamics of quantum correlations under Heisenberg Hamiltonian is explored in section \Romannum{6}. Section \Romannum{7} is devoted for the study under bi-linear bi-quadratic Hamiltonian. The last section explore the conclusion of the work.
\section{Werner state, MEMS and quantum correlation measures}
In this section, we present the description of the Werner state and MEMS with their corresponding density matrices. Further we give the overview of two quantum correlation measures named as squashed entanglement and concurrence, which we have used in current work. 

Werner state\cite{WS} is widely studied quantum state in quantum information theory and it has its great importance in this field. This is a bipartite quantum state in $d\times d$ dimensional Hilbert space which is invariant under all unitary operators and satisfies the equation given below,
\begin{equation}
\rho=(U\otimes U)\rho(U^{\dagger}\otimes U^{\dagger})
\end{equation}
Where $\rho$ is the density matrix of the system. Dealing with two qubits Werner state in $2\times2$ dimensional Hilbert space; the Werner state adopt the the form as given below,
\begin{equation}
\rho_{WS}=\gamma\vert\psi^{-}\rangle\langle\psi^{-}\vert+(1-\gamma)\frac{I}{4}\label{w1}
\end{equation}
Where $\vert\psi^{-}\rangle$ represents a singlet state and I is the $4\times4$ dimensional identity matrix. 

MEMS is an another bipartite quantum state, which has close connection with the bipartite Werner state. The two qubits MEMS was investigated by Munro et al.\cite{MEMS} which is more entangled than two qubit Werner state in terms of concurrence measure and it is experimentally verified also\cite{MEMS2}. The density matrix of two qubits bipartite MEMS can be written as,
\begin{equation}
\rho_{MEMS}=\left[\begin{array}{cccc}
g(\gamma) & 0 & 0 & \frac{\gamma}{2}\\
0 & 1-2g(\gamma) & 0 & 0\\
0 & 0 & 0 & 0\\
\frac{\gamma}{2} & 0 & 0 & g(\gamma)
\end{array}\right]\label{m1}
\end{equation}
This state incorporate a function $g(\gamma)$ with the the following conditions,
\begin{equation}
g(\gamma)=\delta=\left\{\begin{array}{cc}
\frac{1}{3}, &\quad 0\leq\gamma<\frac{2}{3}\\
\frac{\gamma}{2}, &\quad \frac{2}{3}\leq\gamma\leq1
\end{array}\right .\nonumber
\end{equation}
For simplicity throughout this article we consider $g(\gamma)$ as $\delta$. In continuation of the above discussion on quantum states, we would like to bring the attention on quantum correlation measures that we use in our work. Squashed entanglement is a well known quantum correlation measure in quantum information theory. This multipartite entanglement measure satisfy convexity, additivity and super additivity over tensor product in general\cite{sqe2}. Further this measure can also be reduced to the entanglement entropy for pure states. For a quantum state $\rho^{AB}$ defined over the bipartite Hilbert space; squashed entanglement is defined as below,
\begin{equation}
E_{sq}(\rho^{AB})=\textsf{inf}\left\{\frac{1}{2}I(A;B|E):\rho^{ABE}\looparrowright \rho^{AB}\right\}\label{se}
\end{equation}
Here $\looparrowright$ represents the extension of $\rho^{AB}$. Further the infimum will be taken over the set equipped with $\rho^{ABE}$. With the definition of squashed entanglement the following condition is satisfied,
\begin{equation}
\rho^{AB}=Tr_{E}(\rho^{ABE}) 
\end{equation}
The expression $I(A;B|E)$ in the definition of squashed entanglement reads,
\begin{equation}
I(A;B|E)=S(AE)+S(BE)-S(ABE)-S(E)
\end{equation}
Where $S(.)$ represents the Von Neumann entropy in the above expression. Here we mention that the squashed entanglement is invariant under the tripartite exchange permutation symmetry\cite{sqe1}. 

Concurrence\cite{CON} is an another quantum correlation measure which is widely used to determine the entanglement of bipartite quantum systems. Concurrence of a bipartite quantum system of a  density matrix $\rho$ can be defined as,
\begin{equation}
C(\rho)=max\,\{0,\lambda_{1}-\lambda_{2}-\lambda_{3}-\lambda_{4}\}\label{con}
\end{equation}
Where $\lambda_{i}$'s are the decreasing order square root of eigenvalues of $\rho\tilde{\rho}$. 
\begin{equation}
\tilde{\rho}=(\sigma_{y}\otimes\sigma_{y})\rho^{*}(\sigma_{y}\otimes\sigma_{y})
\end{equation}
Where $\tilde{\rho}$ is the result of the spin-flip operation applied on density matrix $\rho$. Here $\sigma_{y}$ is the Pauli Y matrix and $\rho^{*}$ is the complex conjugate of the density matrix $\rho$.

\section{Hamiltonian and Unitary time evolution}
In this section, we present Heisenberg\cite{H1} and bi-linear bi-quadratic Hamiltonian\cite{H2} used for the study. Further we also present the unitary time evolution by using the Schrodinger  time dependent equation.

In this paper we consider Hamiltonians with XXX model in one dimension without magnetic field and following no boundary conditions. Under the above assumption the Heisenberg Hamiltonian reads,
\begin{equation}
H_{1}=-j\sum\limits_{i=1}^N\sigma_i^z\sigma_{i+1}^z\label{h1}
\end{equation} 
Next we consider the non linear extension of the above Hamiltonian named as bi-linear bi-quadratic Hamiltonian. This Hamiltonian is give below,
\begin{equation}
H_{2}=j\sum\limits_{i=1}^N\left[\left(\sigma_i^z\sigma_{i+1}^z\right)-\left(\sigma_i^z\sigma_{i+1}^z\right)^2\right]\label{h2}
\end{equation}
Where $j$ is the coupling constant, $\sigma^{z}$ denotes the Pauli Z matrix and $\left(\sigma_i^z\sigma_{i+1}^z\right)^2$ is the non linear term. These  Hamiltonians are used to investigate the unitary dynamics in bipartite system. 

As per the postulate of quantum mechanics, the unitary time evolution of the physical system is governed by the time dependent Schrodinger equation given as below,
\begin{equation}
i\hbar\frac{\partial }{\partial t}|\psi(t)\rangle=E|\psi(t)\rangle
\end{equation}
Where $E$ is the real energies of the physical system. The solution of this equation is obtained as,
\begin{equation}
|\psi(t)\rangle=e^{\frac{-iHt}{\hbar}}|\psi(0)\rangle 
\end{equation}
The density matrix version of the above equation can be framed as,
\begin{equation}
\rho(t)=U(t).\rho(0).U(t)^{\dagger}\label{tm1}
\end{equation}
Where $U(t)=e^{\frac{-iHt}{\hbar}}$ is the unitary matrix which include the Hamiltonian $H$ in exponential. Here we assume $\hbar=1$ to maintain the simplicity for the present study. This Eq.$\ref{tm1}$ is used in current work to develop the dynamics of quantum correlations in Werner state and MEMS.

\section{Initial state preparation of Werner state and MEMS}
In this section we prepare the initial states according to the requirement of quantum correlations expressed in Eqs.$\ref{se},\ref{con}$.

As per the definition of squashed entanglement we add an additional qubit $(E)$ to the bipartite system; this additional qubit produces the extended density matrix $\rho^{ABE}$. We mention that larger the domain of extension than better the accuracy in squashed entanglement measure. But larger domain of the system makes the calculations of squashed entanglement very difficult, infect it may be considered as NP hard problem\cite{np}\footnote{In computer science the problems whose domain is very large becomes very difficult to solve; such problems are categorized as NP hard.}. In our work we consider the simplest case by considering the auxiliary qubit as external domain for squashed entanglement. To begin with, we express the state vector of additional qubit given as below,
\begin{equation}
|E\rangle=\alpha|0\rangle+\beta|1\rangle\label{e1}
\end{equation}
with
\begin{equation}
|\alpha|^2+|\beta|^2=1\label{x1} 
\end{equation}
The density matrix of the additional qubit $|E\rangle$ is expressed as,
\begin{equation}
\rho_{E}=\left[\begin{array}{cc}
|\alpha|^{2} & \alpha\beta\\
\alpha\beta & |\beta|^{2}
\end{array}\right]
\end{equation}
By using the Eq.$\ref{x1}$, we can rewrite the above density matrix as below, 
\begin{equation}
\rho_{E}=\left[\begin{array}{cc}
|\alpha|^{2} & \alpha\sqrt{1-|\alpha|^{2}}\\
\alpha\sqrt{1-|\alpha|^{2}} & 1-|\alpha|^{2}
\end{array}\right]
\end{equation}
Here we recall that the additional qubit expressed in Eq.$\ref{e1}$ works as an external domain for $\rho^{AB}$ which is the density metrics of bipartite quantum state. Initially the state $\rho^{AB}$ is prepared in Werner State and MEMS respectively, given in equations $\ref{w1},\ref{m1}$. After Heisenberg interaction between qubits $B$ and $E$, the quantum correlations may change in $\rho^{AB}$ with the advancement of time. The density matrices of the initial tripartite system prepared in Werner state and MEMS can be expressed respectively as below,
\begin{equation}
\rho_{W}(0)=\rho_{WS}\otimes\rho_{E}\label{w2}
\end{equation}
and
\begin{equation}
\rho_{M}(0)=\rho_{MEMS}\otimes\rho_{E}\label{m2}
\end{equation}
The above equations will be used for the study in the next sections to explore the dynamics of quantum correlations expressed in Eqs.$\ref{se},\ref{con}$ for Werner state and MEMS. The dynamical study is carried out under two different Hamiltonians given in Eqs. $\ref{h1},\ref{h2}$ in two subsequent sections. In the very next section we explore the mathematical expressions of the dynamics of quantum correlations for quantum states and discuss our results under Heisenberg Hamiltonian. In further section, we investigate the same study under bi-linear bi-quadratic Hamiltonian.

\section{Quantum correlations under Heisenberg Hamilton: The mathematical expressions}
In this section, we investigate the quantum dynamics of quantum correlations in tripartite system presented in Eqs.$\ref{w2}$,$\ref{m2}$. Further we use the time evolution given in Eq.$\ref{tm1}$ under Heisenberg Hamiltonian. Here we present the mathematical expressions obtained for Squashed entanglement and concurrence in Werner states and MEMS in two successive subsections.

\begin{widetext} 
\subsubsection{Squashed entanglement and concurrence for Werner state}
We obtained the time evolution of squashed entanglement for Werner state under Heisenberg Hamiltonian as given below,
\begin{eqnarray}
SqE_{W}(H_{1})&=&0.5+2a^{-1}b\left[(a-G)log(c)+(a+G)log(c^{'})-(a-H)log(d)-(a+H)log(d^{'})\right]\nonumber\\
&&+3b\left[(f)log(0.25f)+(b^{'}+\gamma)log(v)\right]\label{sew1}
\end{eqnarray}
Where,
\begin{eqnarray}
G=\sqrt{\alpha^{2}gg^{'}+a^{2}(1-2\alpha^{2}g)},\quad H=\sqrt{\alpha^{2}f^{'}gg^{'}+a^{2}(1-2\alpha^{2}f^{'}g)},\quad c=(0.5-0.5a^{-1}G),\nonumber
\end{eqnarray}
\begin{eqnarray} 
c^{'}=(0.5+0.5a^{-1}G),\quad d=(0.25-0.25a^{-1}H),\quad d^{'}=(0.25+0.25a^{-1}H)\nonumber
\end{eqnarray}
and,
\begin{eqnarray}
a=e^{2ijt},\quad b=0.180337,\quad b^{'}=0.333333,\quad f=(1-\gamma),\quad f^{'}=(1-\gamma^{2}),\quad f^{''}=(1+\gamma^{*}),\quad f^{'''}=(1+\gamma),\nonumber
\end{eqnarray}
\begin{eqnarray}
f^{''''}=(1-\gamma^{*}),\quad g=(1-\alpha^{2}),\quad g^{'}=(1+a^{4}),\quad h=(1-\delta),\quad h^{'}=(1-2\delta),\quad v=(0.25+0.75\gamma),\nonumber
\end{eqnarray}
\begin{eqnarray}
m=(\gamma-2\delta),\quad m^{'}=(\gamma+2\delta),\quad n=(-0.5\gamma+\delta),\quad n^{'}=(0.5\gamma+\delta).\label{commn}
\end{eqnarray}
The symbols expressed in Eq.$\ref{commn}$ have been used throughout the paper. Further for concurrence we need to obtain the spectrum of eigenvalues with unitary time evolution under Heisenberg Hamiltonian. This spectrum for Werner state is obtained below,
\begin{eqnarray}
\lambda_{W}(H_{1})=\left\{ \frac{1}{4}Xf,\frac{1}{4}Xf,\frac{1}{16}a^{-1}(p-p^{'}),\frac{1}{16}a^{-1}(p+p^{'})\right\}  \label{l1}
\end{eqnarray} 
Where,
\begin{eqnarray}
X=\frac{1}{4}(\alpha^{2}f+gf)^{*},\quad Y=-\frac{1}{2}(a\alpha^{2}\gamma+a^{-1}g\gamma)^{*},\quad p=-4\gamma(Z+Y\alpha^{2}-Z\alpha^{2})-4a^{2}\gamma(Y-Y\alpha^{2}+Z\alpha^{2})+af^{''}f^{'''},\nonumber
\end{eqnarray}
\begin{eqnarray}
Z=-\frac{1}{2}(a^{-1}\alpha^{2}\gamma+ag\gamma)^{*},\quad p^{'}=\sqrt{p^{2}-16\left[4\alpha^{2}\gamma^{2}gg^{'}-a^{2}(1+2\gamma-(3-8\alpha^{2}g)\gamma^{2})\right]\left[YZ-\frac{1}{16}(f^{''})^{2}\right]}\nonumber 
\end{eqnarray} 
and rest of symbols are considered from Eq.$\ref{commn}$.

\subsubsection{Squashed entanglement and concurrence in MEMS}
In this section, we obtain the squashed entanglement for MEMS under unitary time evolution with Heisenberg Hamiltonian. The mathematical expression is obtained as below,  
\begin{eqnarray}
SqE_{M}(H_{1})&=&2a^{-1}b\left[(a-Q)log(k)+(a+Q)log(k^{'})\right]+8b(0.5-\delta)log(h^{'})-(4bh)log(h)-(4b\delta)log(\delta)+\nonumber\\
&&a^{-1}\left[(4bP-2abh)log(l)-(4bP+2abh)log(l^{'})-(4ab\delta)log(\delta)\right]-2b\left[(m)log(n)-(m^{'})log(n^{'})\right] \label{sem1}
\end{eqnarray}
Where,
\begin{eqnarray}
P=\sqrt{a^{2}\left[0.25-(0.5+2\alpha^{2}g)\delta+(0.25+4\alpha^{2}g)\delta^{2}\right]+\alpha^{2}\delta gg^{'}h^{'}},\quad Q=\sqrt{4\alpha^{2}h\delta gg^{'}+a^{2}(1-8\alpha^{2}gh\delta)},\nonumber
\end{eqnarray}
\begin{eqnarray}
k=(0.5-0.5a^{-1}Q),\quad k^{'}=(0.5+0.5a^{-1}Q),\quad l=(0.5-a^{-1}P-0.5\delta),\quad l^{'}=(0.5+a^{-1}P-0.5\delta)\nonumber
\end{eqnarray}
and other symbols taken from Eq.$\ref{commn}$. Further to obtain the concurrence we need to calculate the eigenvalue spectrum; which is given below,
\begin{eqnarray}
\lambda_{M}(H_{1})=\left\{ 0,0,0.25a^{-1}(q-2q^{'}),0.25a^{-1}(q+2q^{'})\right\}  \label{l2}
\end{eqnarray} 
Where,
\begin{eqnarray}
R=0.5\gamma(a\alpha^{2}\gamma+a^{-1}g\gamma)^{*},\quad S=0.5\gamma(a^{-1}\alpha^{2}\gamma+ag\gamma)^{*},\quad q=S+R\alpha^{2}-S\alpha^{2}+a^{2}(R-R\alpha^{2}+S\alpha^{2})+4a\delta\delta^{*}\nonumber
\end{eqnarray}
\begin{eqnarray}
w=\alpha^{2}\gamma^{2}gg^{'}+a^{2}[(1-2\alpha^{2}g)\gamma^{2}-4\delta^{2}],\quad q^{'}=\sqrt{-w(0.5a\alpha^{2}\gamma+0.5a^{-1}g\gamma)^{*}(0.5a^{-1}\alpha^{2}\gamma+0.5ag\gamma)^{*}+w(\delta^{*})^{2}+0.25q^{2}}\nonumber
\end{eqnarray}
and remaining symbols expressed in Eq.$\ref{commn}$.\\
\end{widetext} 

\section{Dynamics of Quantum correlations under Heisenberg Hamiltonian}
In this section, we present the dynamical results of squashed entanglement and concurrence for Werner state and MEMS under Heisenberg Hamiltonian. The dynamics of Squashed entanglement for Werner State and MEMS is governed by the Eqs.$\ref{sew1},\ref{sem1}$ respectively and for Concurrences it is governed by the set of eigenvalues presented in Eqs.$\ref{l1},\ref{l2}$. We note that the
squashed entanglement and concurrence are the functions of the parameters $\alpha,\gamma,j$ and $t$. Further we have shown the comparative study of squashed entanglement and concurrence in graphical results  with varying values of the parameters $(\alpha,\gamma,j,t)$. We divide this study in different cases to understand the dynamical behavior of quantum correlations with respect to different parameters. During the discussion we consider the coupling constant $j$ and time $t$ as a single variable $jt$. The study is explored in five different cases, these cases are given in successive subsections.

\subsubsection*{Case 1: For  the parameter values, $\alpha=0$ and $jt=0$}
\begin{figure}[H]
\centering
\includegraphics[scale=0.86]{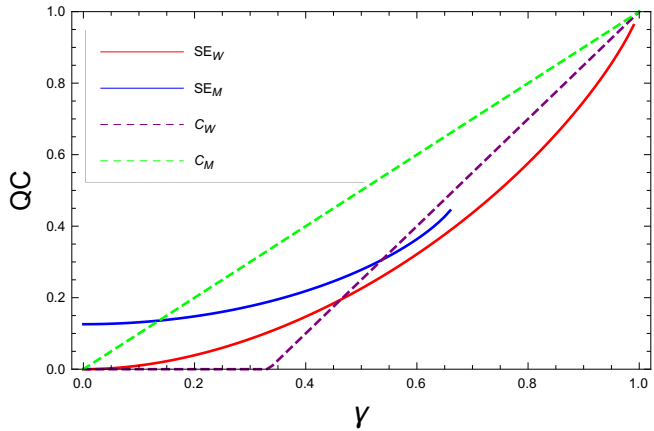} 
\caption{Plot of quantum corrections (QC) vs. $\gamma$}\label{f1}
\end{figure}

In this case, we present the study of quantum correlations in Werner state and MEMS for initial condition i.e. at $\alpha=0,jt=0$. The values of the parameters $(\alpha=0,jt=0)$ maps the tripartite system to the pure bipartite initial state given as in Eqs.$\ref{w1},\ref{m1}$. The initial behavior of quantum correlations with respect to $\gamma$ is shown in Fig.$\ref{f1}$. In this figure $SE_{W}$ represents squashed entanglement of Werner state, $SE_{M}$ corresponds to squashed entanglement of MEMS, $C_{W}$ denotes the concurrence of Werner state and $C_{M}$ represents the concurrence of MEMS which will be followed by the next figures throughout this article.

To proceed the investigation, first we compare the squashed entanglement in both the states; next we compare the concurrence in both the states and finally squashed entanglement vs. concurrence comparison is done.

The squashed entanglement in both the states is plotted in the figure by solid red and blue color. Looking at the graphs we have found that the squashed entanglement in both of the states rise exponentially. The squashed entanglement in Werner state achieve the maximum amplitude as $QC \simeq 1$; while for MEMS it is $QC=0.4428$. It is investigated that,
 MEMS does not exhibit the squashed entanglement after the value of the parameter $\gamma=\frac{2}{3}$. The maximum squashed entanglement amplitude in MEMS is always less than Werner state. At $\gamma=0$, there is no squashed entanglement in Werner state; while for MEMS it has a certain value $QC=0.126365$.

Now we compare the concurrence in both the states which are plotted in figure by dashed purple and green color. Here it is noted that the concurrence grows linearly. The figure shows that the Werner state does not have concurrence between the limit $0\leq \gamma<\frac{1}{3}$; while on the other hand, MEMS has the concurrence for all the range of the parameter $\gamma$. Both the states achieve the maximum amplitude as $QC=1$.

Next, we discuss squashed entanglement and concurrence in both the states. For Werner state, we have compared both the quantum correlations and found that the squashed entanglement is rising exponentially for the range $0\leq \gamma\leq 1$; while the concurrence is increasing in linear fashion between the limit $\frac{1}{3}\leq\gamma\leq 1$. Whereas  concurrence does not exit for the limit $0\leq\gamma<\frac{1}{3}$. It is noted that at the initial value of $\gamma$, the squashed entanglement dominates the concurrence. But after the value of $\gamma=0.4630$ the trade off is reversed i.e. the concurrence dominates the squashed entanglement and achieve the higher amplitude for $0.4630<\gamma\leq 1$. At the value of  $\gamma=0.4630$ both the quantum correlations have the same amplitude and we call this point as Squashed entanglement-Concurrence Equilibrium (SCE) point of Werner state.

For MEMS, we compare both the squashed entanglement and concurrence with interesting trade-off. It is found that the concurrence exists for whole range of $\gamma$; while the squashed entanglement exists in the the range $0\leq\gamma\leq \frac{2}{3}$. Further for the  range of  $ \frac{2}{3}<\gamma\leq 1$ squashed entanglement does not exist. It is also noticed that the SCE point of MEMS exist at $\gamma=0.139$. We also observe that for MEMS with  $0\leq \gamma< 0.139$; the squashed entanglement dominates the concurrence and beyond $\gamma=0.139$, the concurrence dominates the squashed entanglement. 

\subsubsection*{Case 2: For a  fixed value of $\alpha$ and different values of $jt$}
\begin{figure*}
\centering
\includegraphics[scale=0.83]{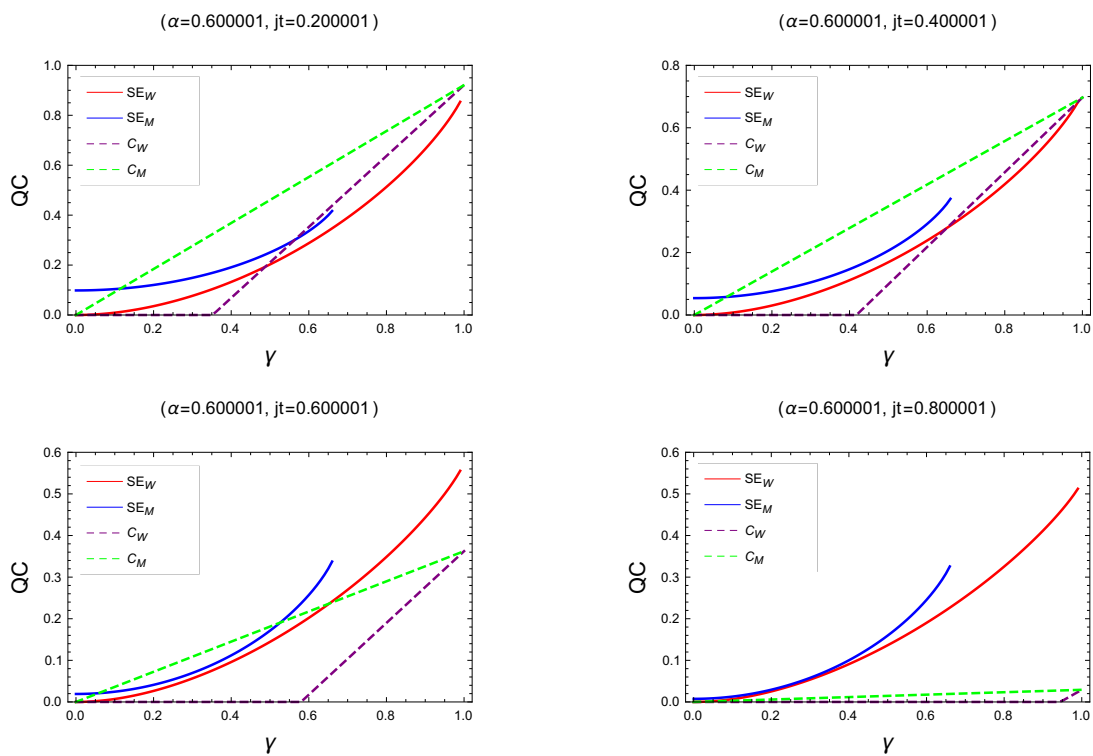}
\caption{Plot of quantum corrections (QC) vs. $\gamma$ with different values of $jt$.}\label{f2}
\end{figure*}

In this case we fix the value of $\alpha$ and plot the graphical results with varying  values of $\gamma$ and $jt$. Few Important results are plotted in Fig.$\ref{f2}$ with $\alpha=0.600001$. 

The graphical plots show that as the value of $jt$ increases the maximum value of all quantum correlations decreases and the rate of amplitude decrement of concurrence is higher in comparison to the amplitude decrement of squashed entanglement.  It is noted that, the sustainability of squashed entanglement in both the state w.r.t the axis of $\gamma$ does not change; it remains same as described in case 1. On the other hand, the sustainability of the concurrence along the axis of $\gamma$ in Werner state changes as the value of the parameter $jt$ increases but for MEMS it does not change. Further we have noticed that for $\gamma=0$, the squashed entanglement of MEMS have certain value, but as time move forward this value decreases and the difference between the squashed entanglement of both the states also decreases.  

\subsubsection*{Case 3: For the  fixed value of $jt$ and different values of $\alpha$}
\begin{figure*}
\centering
\includegraphics[scale=0.83]{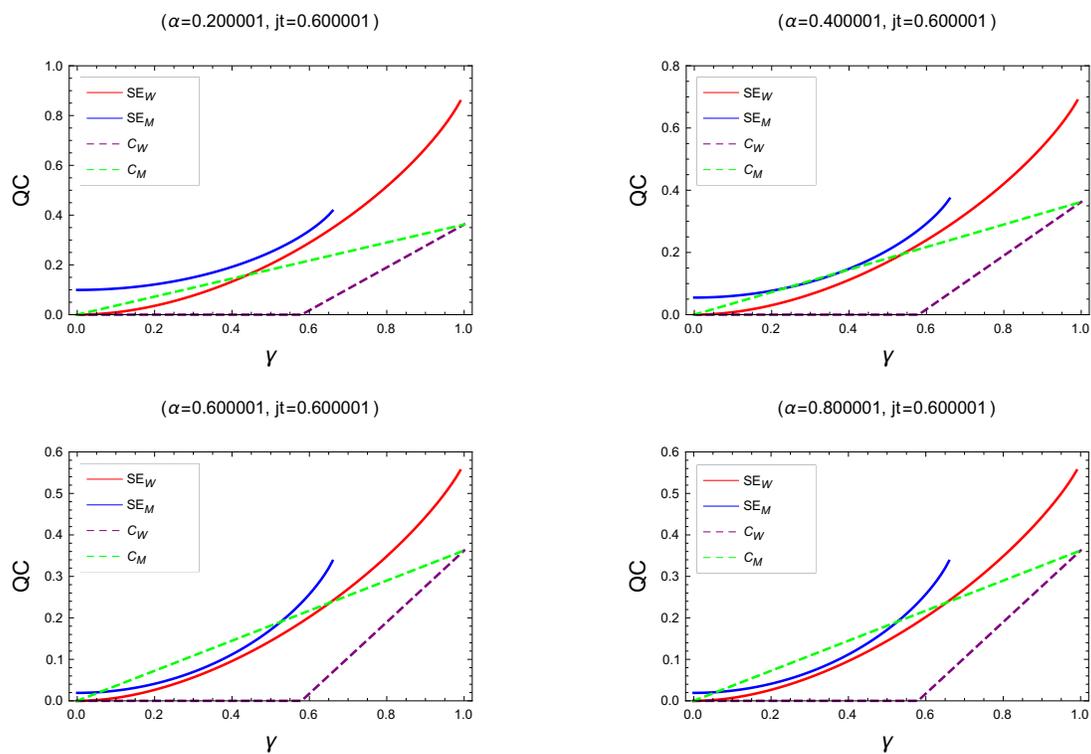}
\caption{Plot of quantum corrections (QC) vs. $\gamma$ with different values of $\alpha$}\label{f3}
\end{figure*}

In this case, we describe the quantum correlations for constant value of $jt$ with respect to varying value of $\alpha$ and $\gamma$. We have shown few of the plots for $jt=0.600001$ and different values of $\alpha$ in Fig.$\ref{f3}$.

Here we notice that as the value of $jt$ is fixed, amplitude of the concurrence of quantum states become freeze at $QC=0.363468$. On the other contrary,  the increment of the value of $\alpha$ decreases the  amplitude of squashed entanglement. It is also noted that the amplitude of squashed entanglement and their gap in both of the states become freeze at $\alpha\geq 0.600001$.

\subsubsection*{Case 4: For the  fixed value of $\alpha$ and different values of $\gamma$}
\begin{figure*}
\centering
\includegraphics[scale=0.83]{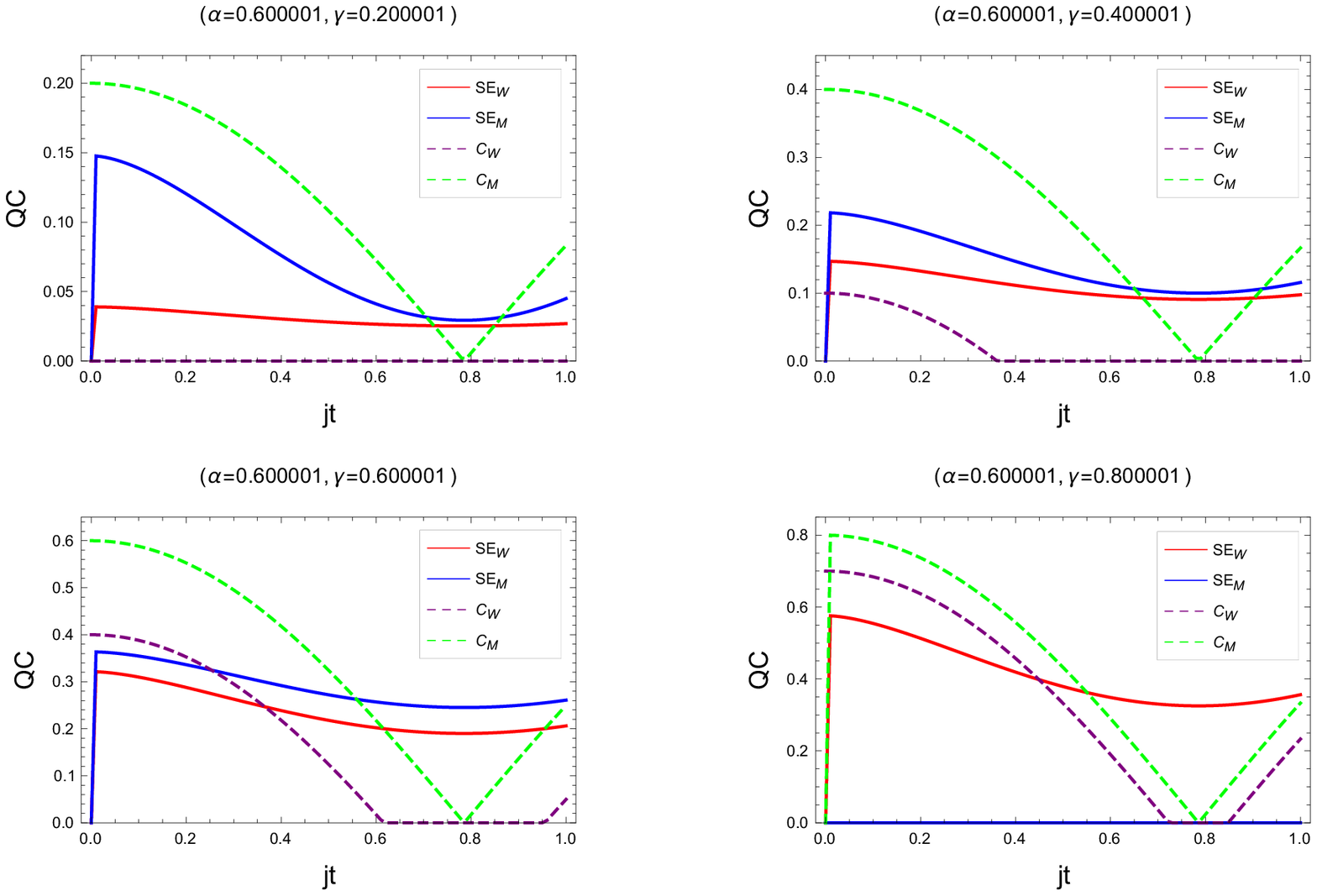}
\caption{Plot of quantum corrections (QC) vs. $jt$ with different values of $\gamma$.}\label{f4}
\end{figure*}

In this section, we explore the study by fixing the value of the parameter $\alpha=0.600001$ with the varying values of $\gamma$ and $jt$. First we discuss the comparison of squashed entanglement in both  the states, next the comparison of concurrence in both of the states is explored. Finally the squashed entanglement and concurrence is compared for each of the states. Here we present the graphical results in Figure $\ref{f4}$.

The squashed entanglement is higher in MEMS than Werner state and amplified in both the states as the value of the parameter $\gamma$ increases. As the value of parameter $jt$ advances the amplitude of squashed entanglement decreases in both the states.

Comparing the concurrence in both the states, we have found that the concurrence is also higher in MEMS than werner states. On the other hand, we have found the important result of Entanglement sudden death (ESD)\cite{sh1,sh2,sh3,sh4,sh5,sh6,sh7,sh8} for the concurrence in Werner state with the increasing  value of $\gamma$. But during the ESD, the concurrence exists in MEMS and this state shows more robust character compared to Werner state. We have found as the value of $\gamma$ increases the ESD zone squeezes.    

Finally we compare the squashed entanglement and concurrence in both the states. We have found at $jt=0$, the  amplitude of saquashed entanglement is always maximum. As the parameter $jt$ advances, both the quantum correlations have decreasing tendency. It is also noted that at $jt=0.787579$, the concurrence of MEMS becomes zero but at this point squashed entanglement exists. For Werner state with $\gamma=0.200001$, the concurrence does not exists, but advancing value of $\gamma$ produces the concurrence in the Werner state with ESD effects. It is important to note that in the absence of concurrence  the squashed entanglement always exist in Werner state. 
\subsubsection*{Case 5: For the  fixed value of $\gamma$ and different values of $\alpha$}
\begin{figure*}
\centering
\includegraphics[scale=0.83]{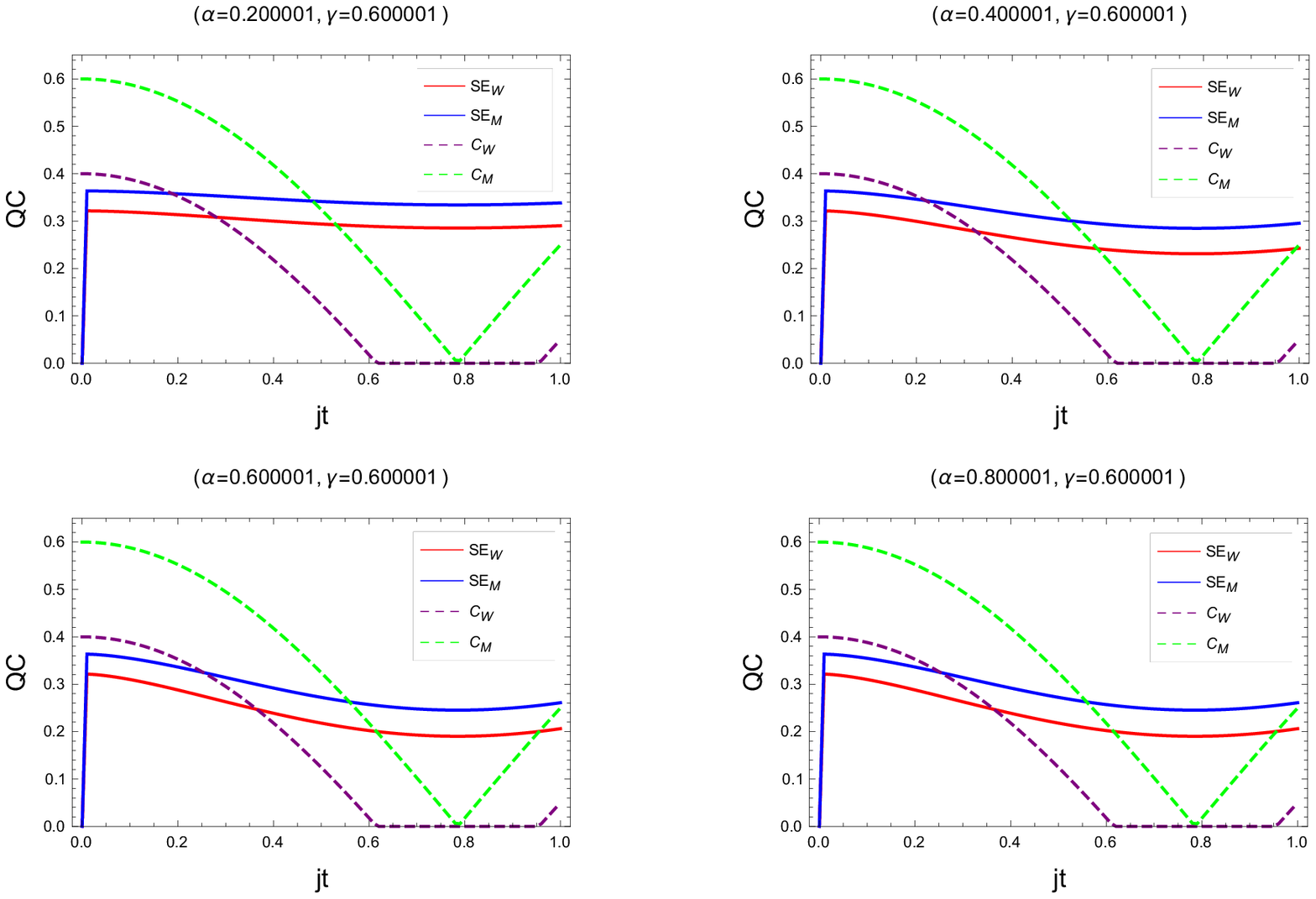}
\caption{Plot of quantum corrections (QC) vs. $jt$ with different values of $\alpha$}\label{f5}
\end{figure*}

In this case, we present the study of the dynamics of quantum correlations by fixing the value of the parameter $\gamma$ as $\gamma=0.600001$ with the varying values of  $\alpha$ and $jt$. The graphical results are shown in Fig.\ref{f5}. 

Here we have found that the maximum amplitude of all the quantum correlations  at $jt=0$ freeze for all values of $\alpha$. We also noticed that the ESD zone of Werner state starts at $jt=0.623712$ and zone width freeze with the advancement of time.\\

\section{Quantum correlations under bi-linear bi-quadratic Hamiltonian: The mathematical expressions}
In this section, we explore the dynamics of quantum correlations in both the states under the bi-linear bi-quadratic Hamiltonian presented in Eq.\ref{h2}. This Hamiltonian incorporates a non-linear term as $(\sigma_{i}^{Z}\sigma_{i+1}^{Z})^{2}$ along with the Heisenberg Hamilton. This term is responsible to cancel out the parameter $jt$ while calculating the time evolution of the system governed by Eq.\ref{tm1}. Hence bi-linear bi-quadratic Hamiltonian preserve the quantum correlations in the system. Here in the following subsections, we present the mathematical functions calculated for  quantum correlations in both the states.  

\begin{widetext} 
\subsubsection{Squashed entanglement and concurrence for Werner state}
The expressions of squashed entanglement and eigenvalue spectrum for concurrence under bi-linear bi-quadratic Hamiltonian with the Eq.\ref{tm1}, is obtained as, 
\begin{eqnarray}
SqE_{W}(H_{2})=1+(3bf)log(0.25f)+b(1+3\gamma)log(v)\label{sew2}
\end{eqnarray} 
and,
\begin{eqnarray}
\lambda_{W}(H_{2})=\left\{ \frac{1}{16}ff^{''''},\frac{1}{16}ff^{''''},\frac{1}{16}ff^{''''},\frac{1}{16}(1+3\gamma)(1+3\gamma^{*})\right\} \label{l3}
\end{eqnarray} 
The required symbols considered in Eq.$\ref{commn}$.\\

We have found, the above equations are independent from the parameters $\alpha$ and $jt$. These equations maps to the Eqs.\ref{sew1} and \ref{l1} with the condition $\alpha=0,jt=0$. Hence the dynamics of quantum correlations in Werner state under bi-linear bi-quadratic Hamiltonian remain same as observed in case 1 with the initial condition.
\subsubsection{Squashed entanglement and concurrence in MEMS}
Here we obtain the expressions of squashed entanglement and eigenvalue spectrum for concurrence under bi-linear bi-quadratic Hamiltonian with Eq.\ref{tm1}. These expressions are give below,
\begin{eqnarray}
SqE_{M}(H_{2})=0.5\left[16b(0.5-\delta)log(h^{'})-(16bh)log(h)-(16b\delta)log(\delta)-4b\left\{ (m)log(n)-(m^{'})log(n^{'})\right\} \right]\label{sem2}
\end{eqnarray}  
and,
\begin{eqnarray}
\lambda_{M}(H_{2})=\left\{ 0,0,(0.25\gamma\gamma^{*}+\delta\delta^{*}-0.5T),0.5(T+0.5\gamma\gamma^{*}+2\delta\delta^{*})\right\} \label{l4}
\end{eqnarray}
Where,
\begin{eqnarray}
T=(\delta\gamma^{*}+\gamma\delta^{*})\nonumber  
\end{eqnarray}
and other symbols expressed in Eq.\ref{commn}\\

Again it is found that the above equations are independent from the parameter $\alpha$ and $jt$; it is happening because of the non-linear term present in the bi-linear bi-quadratic Hamiltonian. The above equations maps to the Eqs.\ref{sem1} and \ref{l2} with the initial condition $\alpha=0,jt=0$. Hence the dynamics of quantum correlations in MEMS remains same as explained in case 1 under the section six with the initial condition. 
\end{widetext} 

\section{Conclusion}
In this article we have studied the unitary dynamics of quantum correlations for Werner state and MEMS under Heisenberg and  bi-linear bi-quadratic Hamiltonian. For initial condition with $\alpha=0,jt=0$, we have found a trade off between squashed entanglement and concurrence in both of the states. With the range $0\leq\gamma<\frac{1}{3}$, the concurrence vanish in Werner state, while squashed entanglement exist. On the other hand, with the range $\frac{2}{3}<\gamma\leq1$ the squashed entanglement vanish in MEMS but concurrence exist for the same range. 

Dealing with Heisenberg Hamiltonian, we have explored this study with varying the parameters $\alpha,\gamma$ and $jt$ in different cases and discuss the impact of every parameters on quantum correlations. We have noticed that the concurrence increases linearly in both  of the states while squashed entanglement increases in exponential fashion. In both  the states, when either one of the parameters $\alpha$ or $jt$ have been increased by keeping another parameter fix; the amplitude of the quantum correlations are decreased. Further the parameter $\gamma$ is responsible to amplify the quantum correlations. We have also investigated the phenomenon of entanglement sudden death (ESD) in Werner state with concurrence measure by varying the value of $\gamma$. But as the value of parameter $\gamma$ is fixed the width of ESD zone is also fixed but in the absence of concurrence the squashed entanglement always exit. More interestingly, it has been observed that the Werner state is more fragile than MEMS in terms of concurrence measure with the parameter $\gamma$.

The most important result we have shown that,  the bi-linear bi-quadratic Hamiltonian does not contribute in the time evolution for both the quantum states and hence it does not disturb the quantum correlations. The non-linear term in this Hamiltonian plays an important role to preserve the quantum correlations and both the states are quite robust under this Hamiltonian.

\end{document}